\documentclass[pre,amsmath,amssymb,twocolumn,superscriptaddress]{revtex4-2}
\usepackage{amsmath,amssymb}
\usepackage[usenames]{color}
\usepackage{amssymb}
\usepackage{dsfont}
\usepackage{grffile}
\usepackage{textcomp}
\usepackage{xspace}
\usepackage{hyperref}
\usepackage{graphicx} 
\usepackage{outlines}
\usepackage{natbib}
\usepackage[dvipsnames]{xcolor} 
\usepackage{ulem}

\newcommand{\tr}{{\rm tr}}

\newcommand{\blue}[1]{{\textcolor{black}{ #1}}}

\newcommand{\sx}{\hat\sigma^x}
\newcommand{\sy}{\hat\sigma^y}
\newcommand{\sz}{\hat\sigma^z}
\newcommand{\sa}{\hat\sigma^a}

\newcommand{\vlr}{v_{\rm LR}}

\newcommand{\sutd}{Science, Mathematics and Technology Cluster, Singapore
University of Technology and Design, 8 Somapah Road, 487372 Singapore}
\newcommand{\sutdepd}{EPD Pillar, Singapore University of Technology and Design, 8 Somapah Road, 487372 Singapore}

\begin{document}
\title{Slow relaxation of out-of-time-ordered correlators in\\ interacting integrable and nonintegrable spin-$\tfrac{1}{2}$ XYZ chains}

\author{Vinitha Balachandran}
\affiliation{\sutd} 
\author{Lea F. Santos}
\affiliation{Department of Physics, University of Connecticut, Storrs, Connecticut 06269, USA}
\author{Marcos Rigol}
\affiliation{Department of Physics, The Pennsylvania State University, University Park, Pennsylvania 16802, USA}

\author{Dario Poletti} 
\affiliation{\sutd} 
\affiliation{\sutdepd} 
\affiliation{The Abdus Salam International Centre for Theoretical Physics, Strada Costiera 11, 34151 Trieste, Italy} 
\affiliation{Centre for Quantum Technologies, National University of Singapore 117543, Singapore}

\date{\today}

\begin{abstract}
  Out-of-time ordered correlators (OTOCs) help characterize the scrambling of quantum information and are usually studied in the context of nonintegrable systems. In this work, we compare the relaxation dynamics of OTOCs in interacting integrable and nonintegrable spin-$\tfrac{1}{2}$ XYZ chains in regimes without a classical counterpart. In both kinds of chains, using the presence of symmetries such as $U(1)$ and supersymmetry, we consider regimes in which the OTOC operators overlap or not with the Hamiltonian. We show that the relaxation of the OTOCs is slow (fast) when there is (there is not) an overlap, independently of whether the chain is integrable or nonintegrable. When slow, we show that the OTOC dynamics follows closely that of the two-point correlators. We study the dynamics of OTOCs using numerical calculations, and gain analytical insights from the properties of the diagonal and of the off-diagonal matrix elements of the corresponding local operators in the energy eigenbasis.
\end{abstract}
  
\maketitle

\setcounter{figure}{0}

\section{Introduction}

Out-of-time-ordered correlators (OTOCs) are a diagnostic of quantum information scrambling~\cite{Witten1998, Maldacena1999, HaydenPreskill2007, SekinoSussking2008, Shenker_2014, SachdevYe1993, Kitaev, LashkariHayden2013,  RobertsStanford2015, CotlerTezuka2017, RobertsSussking2015, HosurYoshida2016, Borgonovi2019,GarciaMataARXIV}, can be used to detect quantum phases \cite{Wang2019,CeresDuan2019, CeresSun2020}, and can be measured experimentally~\cite{LiDu2017, GarttnerRey2017, Wei2018,LandsmanMonroe2019, Niknam2020, JoshiRoos2020, BlokSiddiqi2021, MiYu2021, Jochen2021}.  In quantum systems with a classical limit, OTOCs grow exponentially fast as a result of chaos~\cite{Galitski2017, Hashimoto_2017, Cotler_2018, Ignacio2018,Ray2018, Ch_vez_Carlos_2019, Fortes2019, Rammensee2018, Prakash2020, Bergamasco2019, Rozenbaum2020, Wang2020, wang2020quantum} or of  instabilities in integrable systems~\cite{Pappalardi2018, Hummel2019, Pilatowsky2020, Xu2020, Hashimoto2020,ChavezARXIV}. Due to the relationship with chaos in classical systems, connections between the behavior of OTOCs and the onset of thermalization have been discussed~\cite{HaydenPreskill2007, SekinoSussking2008, Shenker_2014, SachdevYe1993, Kitaev, LashkariHayden2013,  RobertsStanford2015, CotlerTezuka2017, RobertsSussking2015, HosurYoshida2016, Borgonovi2019}. In nonintegrable models with local conserved quantities, OTOCs have been shown to exhibit a slower (algebraic) growth when the operators involved have an overlap with a conserved quantity~\cite{RakovszkyKeyserlingk2018, NahumHaah2017, NahumHaah2018, KeyserlingkSondhi2018, KhemaniHuse2018, Balachandran2021, BalachandranPoletti2022}. This behavior has been explained in terms of the Lieb-Robinson bounds~\cite{Luitz2017, ColmenarezLuitz2020} and of the eigenstate thermalization hypothesis (ETH)~\cite{Balachandran2021, BalachandranPoletti2022}. In this work, we study the unitary dynamics of OTOCs in interacting integrable and nonintegrable spin-$\tfrac{1}{2}$ chains with time-independent Hamiltonians in regimes that do not have a classical counterpart. Our goal is to understand the effect that the overlap of OTOC operators with the Hamiltonian has on the relaxation dynamics of OTOCs.

Interacting integrable systems have been extensively studied theoretically in recent years because of their relevance to experiments~\cite{Bloch_2008, cazalilla_citro_review_11}, and this has resulted in remarkable progress in understanding their dynamics~\cite{calabrese_essler_review_16}. In those systems, after reaching equilibrium following unitary dynamics, observables are not described by traditional Gibbs ensembles~\cite{kinoshita_wenger_06, langen_erne_15, tang_kao_18}, but by generalized Gibbs ensembles that incorporate all the conserved quantities that make the models integrable~\cite{Rigol2007, ilievski_15, vidmar16}. Furthermore, as a result of the presence of an extensive number of conservation laws, the large-distance dynamics of integrable models is different from regular hydrodynamics~\cite{schemmer2019generalized, wilson_malvania_20, malvania_zhang_21}, and it is described by a generalized hydrodynamics~\cite{castro2016emergent, bertini2016transport}.

We compare the time-evolution of OTOCs in interacting integrable and nonintegrable spin-$\tfrac{1}{2}$ XYZ chains. Our analytical insights and numerical results elucidate the role that the overlap between the OTOC operators and the Hamiltonian has on the dynamics of OTOCs. We make those overlaps nonvanishing using symmetries, specifically, $U(1)$ symmetry and supersymmetry. We first highlight that, when the OTOC operators overlap with the Hamiltonian, the OTOC dynamics can mirror the behavior of two-time correlators for long times. We then show that, within the system sizes and timescales that we consider, the dynamics of OTOCs for both interacting integrable and nonintegrable many-body quantum systems can be qualitatively and quantitatively analogous. The OTOCs for interacting integrable and nonintegrable models can exhibit a slow (algebraic-like) relaxation if the OTOC operators overlap with the Hamiltonian, and a faster (exponential-like) relaxation if they do not. Analytical insights are obtained using the relation between the dynamics of the OTOCs and the behavior of the matrix elements of the operators involved in the OTOCs in the energy eigenbasis; specifically, the behavior of the diagonal matrix elements in finite systems with short-range interaction, and the behavior of the off-diagonal elements in the thermodynamic limit.

The presentation is organized as follows. In Sec.~\ref{sec:otoc}, we use the properties of the matrix elements of the operators in the OTOCs, written in the eigenenergy basis, to provide a general understanding of the relaxation dynamics of the OTOCs. The spin-$\tfrac{1}{2}$ XYZ chains in which the OTOCs are studied and the symmetries of their Hamiltonians are discussed in Sec.~\ref{sec:model}. The numerical results are presented in Sec.~\ref{sec:results}. A summary of our results is then provided in Sec.~\ref{sec:conclude}.

\section{Dynamics of OTOCs and Matrix Elements of Operators}
\label{sec:otoc}

\blue{Because of its nontrivial dynamics and universality, we study the infinite-temperature OTOC. For operators $\hat A$ and $\hat B$, it is defined as}
\begin{equation}
\label{otoc}
 O^{AB}(t) = \langle [\hat A(t), \hat B][\hat A(t), \hat B]^\dagger\rangle/2 ,
\end{equation}
where $\hat A(t) = \hat U^{\dagger}(t)\hat A \hat U(t)$ is the time-evolved operator $\hat A$ under the unitary evolution operator $\hat U(t) = \mathcal{T}e^{-i\int_0^t \hat H(\tau) d\tau}$, with $\mathcal{T}$ indicating the time-ordered integration, and $\hat H(t)$ being a general time-dependent Hamiltonian. By infinite-temperature we mean that $\langle\ldots\rangle=\tr(\ldots)/\mathcal{V}$, where $\mathcal{V}$ is the relevant Hilbert space dimension. \blue{In the context of such an average, one says that two operators $\hat A$ and $\hat B$ have a nonzero overlap whenever $\langle \hat A \hat B \rangle \ne 0$.} Equation~(\ref{otoc}) can be rewritten as
\begin{equation}
 O^{AB}(t) =G^{AB}(t)-F^{AB}(t),
\end{equation}
where 
\begin{equation}
 G^{AB}(t)=\langle \hat B \hat A(t) \hat A(t)^\dagger  \hat B^\dagger \rangle
\end{equation}
equals 1 for unitary operators, which are the focus of this work, and
\begin{equation}
 F^{AB}(t)=\langle \hat A(t) \hat B \hat A(t) \hat B\rangle .
 \label{otoc1}
\end{equation}
In what follows, we study the time evolution of $F^{AB}(t)$.

For a time-independent Hamiltonian (our focus here), with eigenenergies $E_\alpha$ and eigenkets $|\alpha \rangle$, Eq.~(\ref{otoc1}) can be written as
\begin{equation}\label{otoc_energybasis}
 F^{AB}(t)= \frac{1}{\mathcal{V}}\sum_{\alpha,\beta,\gamma ,\delta }e^{i(E_\alpha-E_\beta +E_\gamma -E_\delta )t}A_{\alpha \beta}B_{\beta \gamma }A_{\gamma \delta }B_{\delta \alpha} ,
\end{equation}
where $A_{\alpha \beta }=\langle \alpha|\hat A| \beta \rangle$, $B_{\beta \gamma }=\langle \beta |\hat B|\gamma \rangle$. We work in units in which $\hbar=1$.

Next, we discuss two ways in which one can gain an analytic understanding of the OTOCs decay. The first one involves infinite-time  averages in finite systems, and the second one involves the structure of the off-diagonal matrix elements of the operators of interest in the energy eigenbasis in the thermodynamic limit. Our focus is on traceless operators $\hat A$ and $\hat B$ that, in addition to being unitary, are local and Hermitian (i.e., that are physical observables). 

\subsection{OTOC dynamics in the thermodynamic limit and infinite-time averages in finite systems}
\label{sec:otoc_infin} 

Under the assumption of unequal energy spacings,
\begin{equation}
   E_\alpha -E_\beta =E_\gamma -E_\delta \ \ \Longrightarrow \ \
   \begin{cases}
		E_\alpha =E_\beta \text{\ and\ } E_\gamma =E_\delta  \\
		\text{\hspace{1.6cm}or}\\
		E_\alpha =E_\delta \text{\ and\ } E_\beta =E_\gamma \,,
	\end{cases} \label{eq:genspec}
\end{equation}
one finds that the infinite-time average of Eq.~\eqref{otoc_energybasis}, describing the typical long-time results in finite systems of size $L$~\cite{D_Alessio_2016}, is
\begin{align}
\label{eq:otocdiag}
 &F^{AB}_L(\infty) = \frac{1}{\mathcal{V}} \left[\sum_{\alpha}    A_{\alpha \alpha }^2 B_{\alpha \alpha }^2    \right.  \\& \qquad\ \left. +   \sum_{\beta, \alpha \ne \beta}\left(A_{\beta \beta }A_{\alpha \alpha }|B_{\beta \alpha}|^2 + |A_{\beta \alpha}|^2 B_{\beta \beta }B_{\alpha \alpha }\right)\right]. \nonumber
\end{align}

Assumption~\eqref{eq:genspec} for the eigenenergies is traditionally expected to hold for nonintegrable quantum systems~\cite{Srednicki_1998}, and Eq.~\eqref{eq:otocdiag} has been recently verified to be a good approximation for numerical results of OTOCs in such systems~\cite{HuangZhang2019, Balachandran2021}. Assumption~\eqref{eq:genspec} is also expected to hold for interacting integrable quantum systems, which have a Poisson-like level spacing distribution, i.e., the eigenenergies behave as uncorrelated random numbers and, consequently, do not exhibit extensive degeneracies like the ones found in noninteracting models~\cite{Zangara2013}. Here, we show that Eq.~\eqref{eq:otocdiag} can also be used in the context of interacting integrable quantum systems.  

Equation~\eqref{eq:otocdiag} can be further simplified under the assumption that the eigenstate to eigenstate fluctuations of $A_{\alpha \alpha}$ and $B_{\alpha \alpha}$, for {\it all} eigenstates with the same energy density, vanish at least polynomially with the system size~\cite{HuangZhang2019}. Under this assumption, if operator $\hat A$ or $\hat B$ (or both) has (have) a nonzero overlap with the local Hamiltonian (a conserved quantity of the dynamics), then the diagonal matrix elements of the operator decay algebraically with the system size \blue{(we will use symmetries to generate such nonvanishing overlaps).} Since we consider unitary operators, this implies that the first sum in the right-hand side of Eq.~\eqref{eq:otocdiag} decays to zero, as a function of the system size, faster than the second sum. Hence, the results in Ref.~\cite{HuangZhang2019} allow us to simplify Eq.~\eqref{eq:otocdiag} to     
\begin{equation}
\label{eq:simplified}
 F^{AB}_L(\infty) \approx \frac{1}{\mathcal{V}} \sum_{n}\left(A_{\alpha \alpha}^2 + B_{\alpha \alpha}^2\right).
\end{equation}
One can then see that, for $F^{AB}_L(\infty)$ to be nonzero, the diagonal matrix elements of $\hat A$ or $\hat B$ (or both) need to be nonzero. We emphasize that, as mentioned before, a necessary condition for Eq.~\eqref{eq:simplified} to hold is that the operator $\hat A$ or $\hat B$ (or both) has (have) a nonzero overlap with the Hamiltonian. Secondly, for our analyses below, Eq.~\eqref{eq:simplified} is meaningful only when used for sufficiently large but finite systems. In those analyses, Eq.~\eqref{eq:simplified} vanishes identically in the thermodynamic limit.

The assumption that the eigenstate to eigenstate fluctuations of $A_{\alpha \alpha}$ and $B_{\alpha \alpha}$, for {\it all} eigenstates with the same energy density, vanish in the thermodynamic limit has been found to hold for physical observables (represented by few-body operators) in nonintegrable systems~\cite{D_Alessio_2016}, with a decrease of the fluctuations that is exponential with the system size~\cite{Kim_Huse_14, mondaini_fratus_16, LeBlond2019}, as opposed to the weaker polynomial decrease required in Ref.~\cite{HuangZhang2019}. This assumption is violated in integrable systems, but only by a vanishing fraction of the eigenstates with the same energy density. In integrable systems, the variance of the diagonal matrix elements has been found to decay as a power law with the system size~\cite{Biroli_Kollath_10, Ikeda_Watanabe_13, Beugeling2014,Alba_15, zhang_vidmar_22}. This helps understanding why, as we will show, Eq.~\eqref{eq:simplified} can also be used to gain insights into the behavior of OTOCs in interacting integrable systems.

If one invokes the Lieb-Robinson bound, which bounds the speed of the propagation of correlations in local and bounded Hamiltonians~\cite{LiebRobinson1972, CheneauKuhr2012}, then one can use Eqs.~\eqref{eq:otocdiag} and~\eqref{eq:simplified} in finite systems to predict what happens in the thermodynamic limit at finite times. Because of the bound, an accurate description of the evolution of the OTOCs in the thermodynamic limit can be obtained by considering a finite system of size $L_\text{LR}$,
\begin{equation}
F^{AB}_{L=\infty}(t) \approx F^{AB}_{L_\text{LR}}(t), \label{eq:inf_to_finite_size}
\end{equation} 
where $L_\text{LR}\equiv s\;\vlr \;t$, $\vlr$ is the Lieb-Robinson velocity, and $s$ is a real number larger than $1$. 

Assuming that the system is maximally scrambled within the region of size $L_\text{LR}$, one can write 
\begin{equation}
F^{AB}_{L=\infty}(t) \approx F^{AB}_{L_\text{LR}}(\infty), \label{eq:inf_time}
\end{equation}
which, since $L_\text{LR}$ increases with time, is a time-dependent quantity. If $F^{AB}_{L}(\infty)$ decays algebraically with the system size, {\it i.e.}, if
\begin{equation}
F^{AB}_{L}(\infty)\propto \frac{1}{L^{\eta}}, \label{eq:polyn_decay}
\end{equation}
which, as mentioned before, occurs if the operator $\hat A$ or $\hat B$ has a nonzero overlap with the local Hamiltonian~\cite{HuangZhang2019}, then the OTOC in the thermodynamic limit decays algebraically in time
\begin{equation}
F^{AB}_{L=\infty}(t) \propto \frac{1}{t^\eta}. \label{eq:slow_decay}
\end{equation}

If the system is not maximally scrambled as assumed before, then the decay of $F^{AB}_{L=\infty}(t)$ is slower, i.e., $F^{AB}_{L_\text{LR}}(\infty)$ is a lower bound for the relaxation dynamics of $F^{AB}_{L=\infty}(t)$. Slower dynamics than the one predicted by this bound occurs, for example, in systems that undergo prethermalization~\cite{LuitzKhemani, Lee2019}. Also, if the diagonal matrix elements of the traceless operators $\hat A$ and $\hat B$ are already vanishingly small in finite systems, then the infinite-time average $F^{AB}_{L}(\infty) \approx 0$, and the decay of $F^{AB}_{L=\infty}(t)$ can be faster than algebraic, e.g., exponential~\cite{RakovszkyKeyserlingk2018, Balachandran2021}. A comprehensive analysis of these features for nonintegrable systems can be found in Ref.~\cite{Balachandran2021}.

We stress that, as we will show using numerical calculations in the following sections, the results in this section can be invoked in the context of both interacting integrable and nonintegrable many-body systems. Both classes of systems exhibit ``generic spectra'' in the sense of Eq.~\eqref{eq:genspec} and a vanishing fraction of eigenstates that do not exhibit eigenstate thermalization for (few-body) operators representing physical observables.

\subsection{Slow dynamics of OTOCs and two-time correlators}
\label{sec:otoc_var}

One can also gain an understanding of the decay of the OTOCs in interacting integrable and nonintegrable many-body systems using the properties of the off-diagonal matrix elements of local operators in the energy eigenstates. To do this, we note that Eq.~(\ref{eq:simplified}) corresponds to the infinite-time average of the two-time correlator
\begin{equation}
C^{AB}(t) = \langle \hat A(t)\hat A(0) \rangle + \langle \hat B(t)\hat B(0) \rangle. \label{eq:C_AB}     
\end{equation}
Within the Lieb-Robinson argument discussed earlier in the context of Eqs.~\eqref{eq:inf_to_finite_size}--\eqref{eq:slow_decay}, this implies that the slow dynamics of the OTOCs is a result of the slow dynamics of the two-time correlator $\langle \hat A(t) \hat A(0) \rangle$ or $\langle \hat B(t) \hat B(0) \rangle$. The OTOCs can decay fast in time only if both two-time correlators decay fast in time.

Let us then analyze the dynamics at long times of a two-time correlator, say of operator $\hat A$, to see how it is related to the structure of its off-diagonal elements in the eigenenergy basis. In nonintegrable systems, one can use that, according to the ETH, the matrix elements of local operators in the energy eigenbasis can be written as~\cite{Deutsch, Srednicki, Rigol2008, D_Alessio_2016}
\begin{equation}
A_{\alpha\beta} = A(\bar{E}_{\alpha\beta})\delta_{\alpha\beta} + e^{-S(\bar{E}_{\alpha\beta})/2} f^A(\bar{E}_{\alpha\beta}, \omega_{\alpha\beta}) R_{\alpha\beta} ,
\label{eth}
\end{equation}
where $\bar{E}_{\alpha\beta}=(E_\alpha+E_\beta)/2$, $\omega_{\alpha\beta} = E_\alpha - E_\beta$, $S(\bar{E})$ is the thermodynamic entropy at energy $\bar{E}$, the functions $A(\bar{E}_{\alpha\beta})$ and $f^A(\bar{E}_{\alpha\beta}, \omega_{\alpha\beta})$ are smooth functions of their arguments, and $R_{\alpha\beta}$ are random numbers with zero mean and unit variance. For bounded lattice Hamiltonians like the ones of interest here, the overwhelming majority of the eigenstates is at ``infinite temperature'', namely, the extensive part of their eigenenergies is $E_\infty\equiv\tr(\hat H)/\mathcal{V}$. When $\bar{E}_{\alpha\beta}=E_\infty$, $\exp[-S(E_\infty)]\simeq 1/\mathcal{V}$, and the ETH ansatz simplifies to
\begin{equation}
A_{\alpha\beta}\simeq  A(E_\infty)\delta_{\alpha\beta} + \frac{1}{\sqrt{\mathcal{V}}} f^A(E_\infty, \omega_{\alpha\beta}) R_{\alpha\beta}.
\label{eth_inf}
\end{equation}
With that in mind, one can write the two-time correlator for the operator $\hat A$ as 
\begin{equation}\label{twotime_tevol}
 \langle A(t)A(0) \rangle \simeq \frac{1}{\mathcal{V}^2}\sum_{\alpha, \beta}e^{i \omega_{\alpha\beta} t} |R_{\alpha\beta}|^2 \left|f^A(E_\infty, \omega_{\alpha\beta})\right|^2.
\end{equation}
Since the $f^A$ functions decay rapidly with increasing frequency~\cite{D_Alessio_2016}, one can replace the sums by integrals using that the density of states at infinite temperature is $\simeq \mathcal{V}$, and we find 
\begin{equation}\label{twotime_tevol_final}
 \langle A(t)A(0) \rangle \simeq  \int d\omega \cos (\omega t) \left| f^A(E_\infty, \omega)\right|^2.   
\end{equation}
Hence, the low-frequency behavior of $\left|f^A(E_\infty, \omega)\right|^2$ (or $\left|f^B(E_\infty, \omega)\right|^2$) determines the relaxation dynamics of the OTOCs. Specifically, structure (nonzero $\omega$ derivative) at low frequency results in a slow decay of the OTOCs while lack of structure (e.g., a plateau) results in a fast (exponential-like) decay of the OTOCs.

It has been recently shown that, like in nonintegrable many-body quantum systems, the vanishing off-diagonal matrix elements of local operators in the energy eigenstates of interacting integrable many-body quantum systems are measure zero of all the matrix elements, i.e., the off-diagonal matrix elements are dense~\cite{LeBlond2019}. This is to be contrasted to what happens in noninteracting systems, in which the vanishing matrix elements are measure one, i.e., the off-diagonal matrix elements are sparse~\cite{zhang_vidmar_22}. Hence, in interacting integrable many-body quantum systems one can define a meaningful function that parallels $|f^{A/B}(\bar E,\omega)|^2$ in Eq.~\eqref{eth}. At $\bar E=E_\infty$, such a function has been shown to be a smooth function of $\omega$ for various local observables~\cite{LeBlond2019, LeBlond2020, Brenes_LeBlond_20, brenes_goold_20}. Consequently, we expect that the analysis leading to Eq.~\eqref{twotime_tevol_final} holds for interacting integrable systems and, with it, the analytic insights that the equation provides. We should add that an important difference between the behavior of the off-diagonal matrix elements of nonintegrable and interacting integrable systems, which does not affect our analysis here, is that in the former the $R_{\alpha\beta}$ random numbers are normally distributed while in the latter they are not~\cite{LeBlond2019}. For integrable systems it was recently shown that the distributions are well described by generalized Gamma distributions~\cite{zhang_vidmar_22}.

\section{Hamiltonian and symmetries}
\label{sec:model}

We consider the spin-$\tfrac{1}{2}$ XYZ model described by the following Hamiltonian,
\begin{align}
\hat H & = \hat H_{I}+ \hat H_{NI}, 
\label{ham2} \\
\hat H_{I} & = \sum_{l=1}^{L-1}\left[J^x \sx_l\sx_{l+1} + J^y\sy_l\sy_{l+1}  + J^z \sz_l\sz_{l+1}\right], 
\nonumber
\\
\hat H_{NI} & = \Lambda \sum_{l=1}^{L-1}(-1)^l \sz_l\sz_{l+1}. \nonumber 
\end{align}
where $J^a$ are the coupling strengths along the $a=x$, $y$, and $z$ directions, $\sa_l$ is represented by the $a$ Pauli matrix for site $l$, and $L$ is the number of lattice sites. $\hat H_{I}$ is an interacting integrable Hamiltonian that can be solved analytically using the eight vertex model~\cite{Baxter1971, BAXTER1972193}. The addition of the staggered interactions, $\hat H_{NI}$, of magnitude $\Lambda$ along the $z$-direction, breaks integrability.  

The symmetries of the total Hamiltonian $\hat H$ are:

(i) The parity operators along each axis, $\hat P^a = \prod_l \sa_l$, commute with the total Hamiltonian. (ia) When $L$ is even, the parity operators also commute with each other. As a result, the diagonal matrix elements of the local operators $\sa_l$ in the energy eigenstates are exactly zero (numerically, we find them to be of the order of $10^{-14}$), unless there are other symmetries such as those discussed next in (ii) and (iii). We can understand this as follows: For $L$ even, the energy eigenkets $|\alpha\rangle$ (where $\hat H|\alpha\rangle=E_\alpha|\alpha\rangle$) are simultaneous eigenkets of the three operators $\hat P^a$. Hence, we can write 
\begin{equation}
    \langle \alpha | \sz_l | \alpha \rangle = \langle \alpha | \hat P^x \left( \hat P^x\sz_l \hat P^x \right) \hat P^x | \alpha \rangle = - \langle \alpha | \sz_l | \alpha \rangle=0, \label{eq:even_zero} 
\end{equation} 
where we used that $\hat P^x\sz_l \hat P^x= - \sz_l$ and that $\hat P^x | \alpha \rangle = \pm | \alpha \rangle$. (ib) When the system size $L$ is odd, the parity operators anticommute with each other. As a result, one finds that the diagonal matrix elements of the local operators $\sa_l$ in the energy eigenstates need not vanish, which is what we observe numerically. We can understand this as follows: For $L$ odd, the energy eigenkets $|\alpha\rangle$ are not simultaneous eigenkets of the three operators $\hat P^a$. Let us choose the energy eigenkets to be simultaneous eigenkets of $\hat P^z$. If $|\alpha_+\rangle$ is an energy eigenket with eigenenergy $E_\alpha$ satisfying $\hat P^z|\alpha_+\rangle=|\alpha_+\rangle$, then $\hat P^x |\alpha_+\rangle = |\alpha_-\rangle$ is an energy eigenket with eigenenergy $E_\alpha$ (because $\hat P^x$ and $\hat H$ commute) satisfying $\hat P^z|\alpha_-\rangle=-|\alpha_-\rangle$ (because $\hat P^x$ and $\hat P^z$ anticommute), i.e., the energy spectrum is doubly degenerate~\cite{Giampaolo_2019}. For the diagonal matrix elements of the local operators, instead of Eq.~\eqref{eq:even_zero} we have
\begin{align}
    \langle \alpha_+ | \sz_l | \alpha_+ \rangle & = \langle \alpha_+ | \hat P^x \left( \hat P^x\sz_l \hat P^x \right) \hat P^x | \alpha_+ \rangle \nonumber\\ & = - \langle \alpha_- | \sz_l | \alpha_- \rangle,  
\end{align}
i.e., they need not vanish. With increasing system size, one expects that the difference between the results obtained for chains with $L$ and $L+1$ sites decreases, and vanish in the thermodynamic limit. Our numerical results are consistent with that expectation.

(ii) In general, the integrable Hamiltonian $\hat H_I$ does not conserve total magnetization in any direction, except when two of the coupling parameters are equal, in which case the model exhibits a $U(1)$ symmetry and reduces to the spin-$\tfrac{1}{2}$ XXZ chain (upon relabeling, if needed, $x$, $y$, and $z$ so that $J^z\neq J^x=J^y$). For $\Lambda\neq0$, the $U(1)$ symmetry is only present for $\hat H$ if $J^x=J^y$.

One can show that if $\hat H$ has $U(1)$ symmetry, i.e., if $\hat{{\cal S}}^z=\sum_l \hat{\sigma}^z_l$ is conserved, then the diagonal matrix elements of the local operators $\sz_l$ in the energy eigenstates need not vanish. Considering $L$ even, we have already mentioned that $\hat{P}^x$, $\hat{P}^y$, and $\hat{P}^z$ commute with each other and with the Hamiltonian. However, only $\hat{P}^z$ commutes with $\hat{{\cal S}}^z=\sum_l \hat{\sigma}^z_l$. If we choose the energy eigenstates to be simultaneous eigenstates of $\hat{{\cal S}}^z$, then we have 
\begin{align}
    \langle \alpha | \sz_l | \alpha \rangle & = \left( \langle \alpha | \hat P^x \right)\left( \hat P^x\sz_l \hat P^x \right) \left(\hat{P}^x | \alpha \rangle \right)\nonumber\\ & = - \left(\langle \alpha | \hat P^x \right) \sz_l  \left(\hat P^x | \alpha \rangle \right).
\end{align}
Since $\hat{P}^x$ and $\hat{{\cal S}}^z$ do not commute, $\hat P^x | \alpha \rangle$ does not need to be $e^{i \phi} |\alpha \rangle$, so $\langle \alpha|\hat{\sigma}^z_l |\alpha \rangle$ does not need to vanish. However, one can show that the matrix elements $\langle \alpha|\hat{\sigma}^x_l |\alpha \rangle$ must vanish
\begin{equation}
    \langle \alpha | \sx_l | \alpha \rangle  =  \langle \alpha | \hat P^z \left( \hat P^z\sx_l \hat P^z \right) \hat{P}^z | \alpha \rangle= - \langle \alpha | \sx_l  | \alpha \rangle=0.
\end{equation}

(iii) The integrable Hamiltonian $\hat H_{I}$ exhibits supersymmetry when~\cite{hagendorf2012eight, lienardy2020integrable}
\begin{equation}
    J^xJ^y+J^xJ^z+J^yJ^z=0. \label{eq:supersymmetry}
\end{equation} 
Defining the operator $\hat P^a_e = \prod_{l=1}^{L/2}\hat \sigma^a_{2l}$, which acts only on even sites, we get $\hat P^z_e \hat H_I (J^x, J^y, J^z) \hat P^z_e = \hat H_I (-J^x, -J^y, J^z)$, and similarly for $a=x,y$. Hence, the supersymmetry also holds for 
\begin{equation}
J^xJ^y \pm J^xJ^z \pm J^yJ^z=0. 
\label{eq:supersymmetryB}
\end{equation} 

The previous symmetry analysis shows that the $\sa_l$ (where $a=x$, $y$, or $z$) operators have vanishing diagonal matrix elements, i.e., a vanishing overlap with the Hamiltonian~\eqref{ham2}, unless the Hamiltonian has special symmetries such as $U(1)$ or supersymmetry. This means that, by tuning the Hamiltonian parameters to be at those special symmetry points (or away from them), we can test the effect that the overlap (or lack thereof) of operators with the Hamiltonian has on the relaxation dynamics of the OTOCs at and away from integrability.

Throughout this paper, we fix $J^y=1.0$ to be our energy scale, $J^z=1.5$, and scan across $J^x$. For the integrable chain $\Lambda=0$, and for the nonintegrable one we choose $\Lambda=0.2$.\\

\section{Numerical Results}
\label{sec:results}

The fact that the diagonal matrix elements of $\hat \sigma^{a}_l$ are vanishing away from the points at which the Hamiltonian of the spin-$\tfrac{1}{2}$ XYZ chain exhibits $U(1)$ symmetry or supersymmetry, and can be nonvanishing as one approaches those points, is illustrated in Figs.~\ref{fig:fig1}(a) and~\ref{fig:fig1}(b). There we plot the \blue{average} (over the entire energy spectrum) of the absolute values of the diagonal matrix elements of $\sz_{L/2}$ and $\sx_{L/2}$, respectively, as functions of $J^x$ for $L=14$. The \blue{averages} are labeled as 
\blue{\begin{equation}
 S^z=\frac{1}{\mathcal{V}}\sum_{\alpha}| (\sz_{L/2})_{\alpha\alpha}|, 
\end{equation}}
and
\blue{\begin{equation}
 S^x=\frac{1}{\mathcal{V}}\sum_{\alpha}| (\sx_{L/2})_{\alpha\alpha}|.
\end{equation}}
When $J^x=1$, since $J^x=J^y$, the Hamiltonian has $U(1)$ symmetry and conserves total magnetization in the $z$-direction. For values of $J^x$ close to 1 the total magnetization is a nearly-conserved quantity, which is known to result in prethermal behavior both in the integrable and nonintegrable cases~\cite{Mallayya2019}. This explains the ``broad'' peak of $S^z$ at $J^x=1$ in Fig.~\ref{fig:fig1}(a) for both the integrable and the nonintegrable chains. \blue{In the discussions that follow, we refer to the $J^x=1.0$, $\Lambda=0$ point as the ``$U(1)$ integrable point'' and to the $J^x=1.0$, $\Lambda=0.2$ point as the ``$U(1)$ nonintegrable point.''} The sharper peaks at $J^x=0.6$ for both $S^z$ in Fig.~\ref{fig:fig1}(a) and $S^x$ in Fig.~\ref{fig:fig1}(b) occur only in the integrable chain and are caused by the supersymmetric point satisfying Eq.~\eqref{eq:supersymmetryB}. \blue{In the discussions that follow, we refer to $J^x=0.6$, $\Lambda=0$ point as the ``SUSY point.''}

The results in the main panels in Figs.~\ref{fig:fig1}(a) and~\ref{fig:fig1}(b) are for a chain in which $L$ is even. For an odd number of sites, as mentioned before, the diagonal matrix elements of $\hat \sigma^{a}_l$ need not vanish even if they vanish for an even number of sites. The inset in Fig.~\ref{fig:fig1}(a) [Fig.~\ref{fig:fig1}(b)] shows $S^z$ [$S^x$] as a function of $L$ for $J^x=1,\; 1.6$ and $\Lambda=0,\;0.2$ [$J^x=0.6 ,\; 1.3$ and $\Lambda=0, \;0.2$]. \blue{At the $U(1)$ integrable and nonintegrable points in the inset of Fig.~\ref{fig:fig1}(a), and at the SUSY point in the inset of Fig.~\ref{fig:fig1}(b),} both the $L$ even and $L$ odd systems have nonzero diagonal values. For the other parameters we only show results for $L$ odd. In the latter cases we see that $S^z$ and $S^x$ decrease rapidly as $L$ (odd) increases, i.e., the results for chains with $L$ odd approach the vanishing results for $L$ even. Hence, we expect that Fig.~\ref{fig:fig1}(a) and Fig.~\ref{fig:fig1}(b) are representative of what happens in large chains no matter whether $L$ is even or odd. (See Appendix~\ref{app:odd-even} for a comparison between the dynamics of OTOCs in systems with even and odd numbers of lattice sites.)

Our discussion thus far allows us to conclude that the peaks in Figs.~\ref{fig:fig1}(a) and~\ref{fig:fig1}(b) establish the Hamiltonian parameters for which the corresponding operators exhibit an overlap with the Hamiltonian. Next, we study how such an overlap (or lack thereof) affects the time evolution of the OTOCs involving those operators.

\subsection{Relaxation dynamics and diagonal matrix elements of local operators}
\label{sec:numdiag}

\begin{figure*}[!t]
\includegraphics[width=0.8\textwidth]{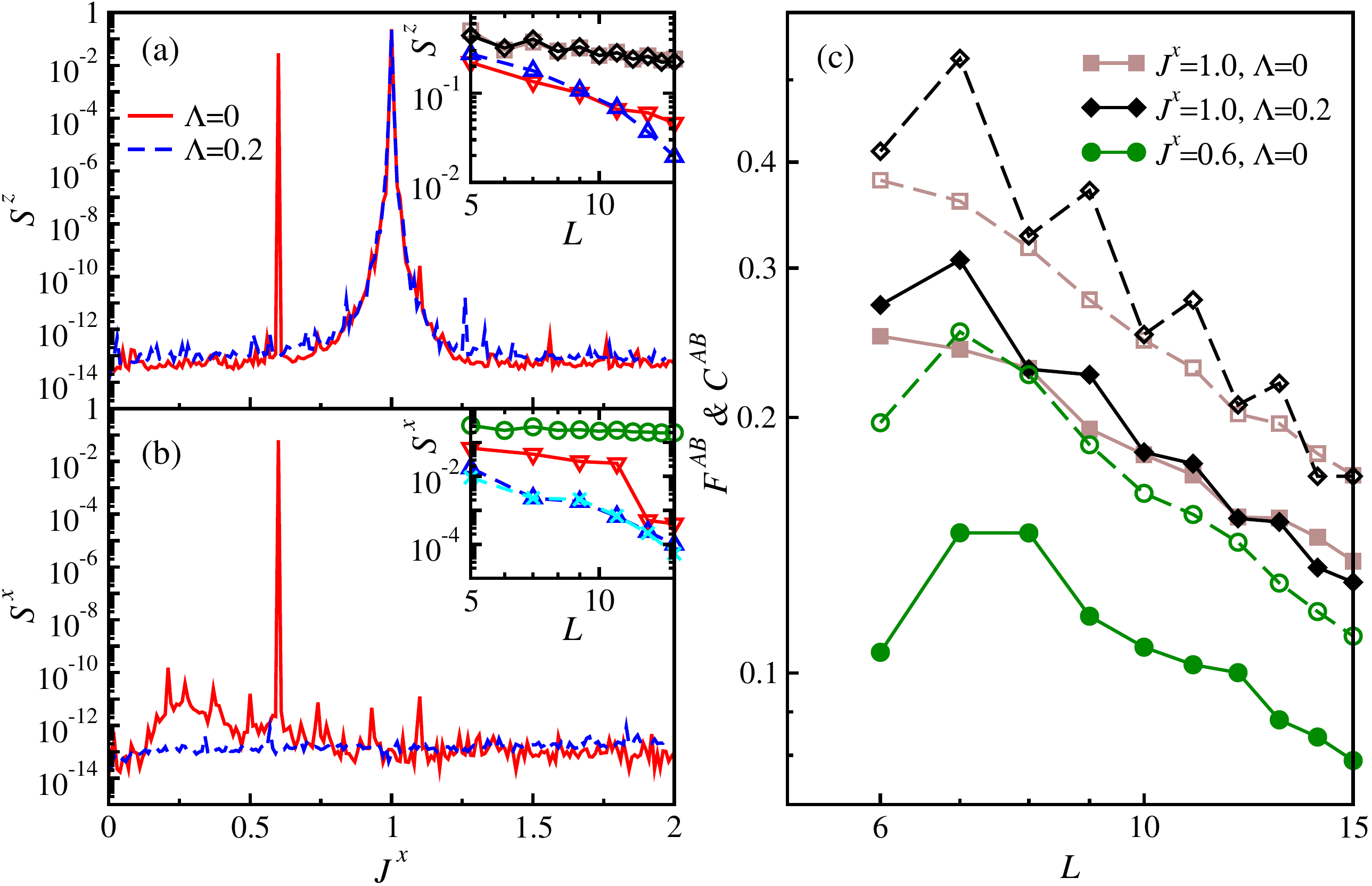}
\vspace{-0.2cm}
\caption{\blue{Average} of the absolute values of the diagonal matrix elements of $\sa_{L/2}$ in the energy eigenbasis. (a) \blue{$S^z=\sum_{\alpha}| (\sz_{L/2})_{\alpha\alpha}|/\mathcal{V}$} and (b) \blue{$S^x=\sum_{\alpha}| (\sx_{L/2})_{\alpha\alpha}|/\mathcal{V}$}, plotted as functions of the coupling parameter $J^x$ for a chain with an even number of spins ($L=14$) for the integrable, $\Lambda=0$, and nonintegrable, $\Lambda=0.2$, cases. Inset in panel (a), scaling of $S^z$ as a function of $L$ for $J^x=1.0,\,\Lambda=0$ [\blue{$U(1)$ integrable point}, \textcolor{brown}{$\square$}] and $J^x=1.0,\,\Lambda=0.2$ [\blue{$U(1)$ nonintegrable point}, $\Diamond$]; and (only showing results for odd values of $L$) for $J^x=1.6,\,\Lambda=0$ (\textcolor{red}{$\bigtriangledown$}) and $J^x=1.6,\,\Lambda=0.2$ (\textcolor{blue}{$\bigtriangleup$}). Inset in panel (b), scaling of $S^x$ as a function of $L$ for $J^x=0.6,\,\Lambda=0$ (\blue{SUSY point}, \textcolor{green}{$\bigcirc$}); and (only showing results for odd values of $L$) for $J^x=0.6,\,\Lambda=0.2$ (\textcolor{cyan}{$\times$}), $J^x=1.3,\,\Lambda=0$ (\textcolor{red}{$\bigtriangledown$}), and $J^x=1.3,\Lambda=0.2$ (\textcolor{blue}{$\bigtriangleup$}). (c) Infinite-time average $F^{AB}(\infty)$ (solid  lines with filled symbols) and $C^{AB}(\infty)$ (dashed lines with empty symbols) as functions of $L$ for $\hat A=\sz_{L/2-1}$ and $\hat B=\sz_{L/2+2}$ (\textcolor{brown}{$\square$} for $\Lambda=0$ and $\Diamond$ for $\Lambda=0.2$) and for $\hat A=\sx_{L/2-1}$ and $\hat B=\sx_{L/2+2}$ (\textcolor{green}{$\bigcirc$}). In all panels, we show results for the XYZ chain with $J^y=1.0$ and $J^z=1.5$.}
\label{fig:fig1}
\end{figure*}

For the three special sets of parameters for which large peaks are seen in Fig.~\ref{fig:fig1}(a), namely, \blue{the $U(1)$ integrable and nonintegrable points and the SUSY point}, as well as in Fig.~\ref{fig:fig1}(b) \blue{at the SUSY point}, the \blue{average} of the absolute values of the diagonal matrix elements decreases polynomially with $L$, as can be inferred from the slow decay of $S^z$ in the inset of Fig.~\ref{fig:fig1}(a) and of $S^x$ in the inset of Fig.~\ref{fig:fig1}(b). This leads to the algebraic decay of the infinite-time averages of the OTOCs [two-time correlators] as functions of the system size, as predicted by Eq.~\eqref{eq:polyn_decay} [Eq.~\eqref{eq:otocdiag}] and confirmed numerically in Fig.~\ref{fig:fig1}(c). In Fig.~\ref{fig:fig1}(c), we plot $F^{\sz_{L/2-1} \sz_{L/2+2}}(\infty)$ [$C^{\sz_{L/2-1} \sz_{L/2+2}}(\infty)$] for both the integrable and the nonintegrable chains with \blue{$U(1)$ symmetry}, and $F^{\sx_{L/2-1} \sx_{L/2+2}}(\infty)$ [$C^{\sx_{L/2-1} \sx_{L/2+2}}(\infty)$] for the integrable chain with \blue{supersymmetry}. According to Eq.~\eqref{eq:slow_decay}, we therefore expect that in the thermodynamic limit the relaxation dynamics of the OTOCs for these sets of parameters exhibit an algebraic decay in time.

\begin{figure}[!t]
\includegraphics[width=0.98\columnwidth]{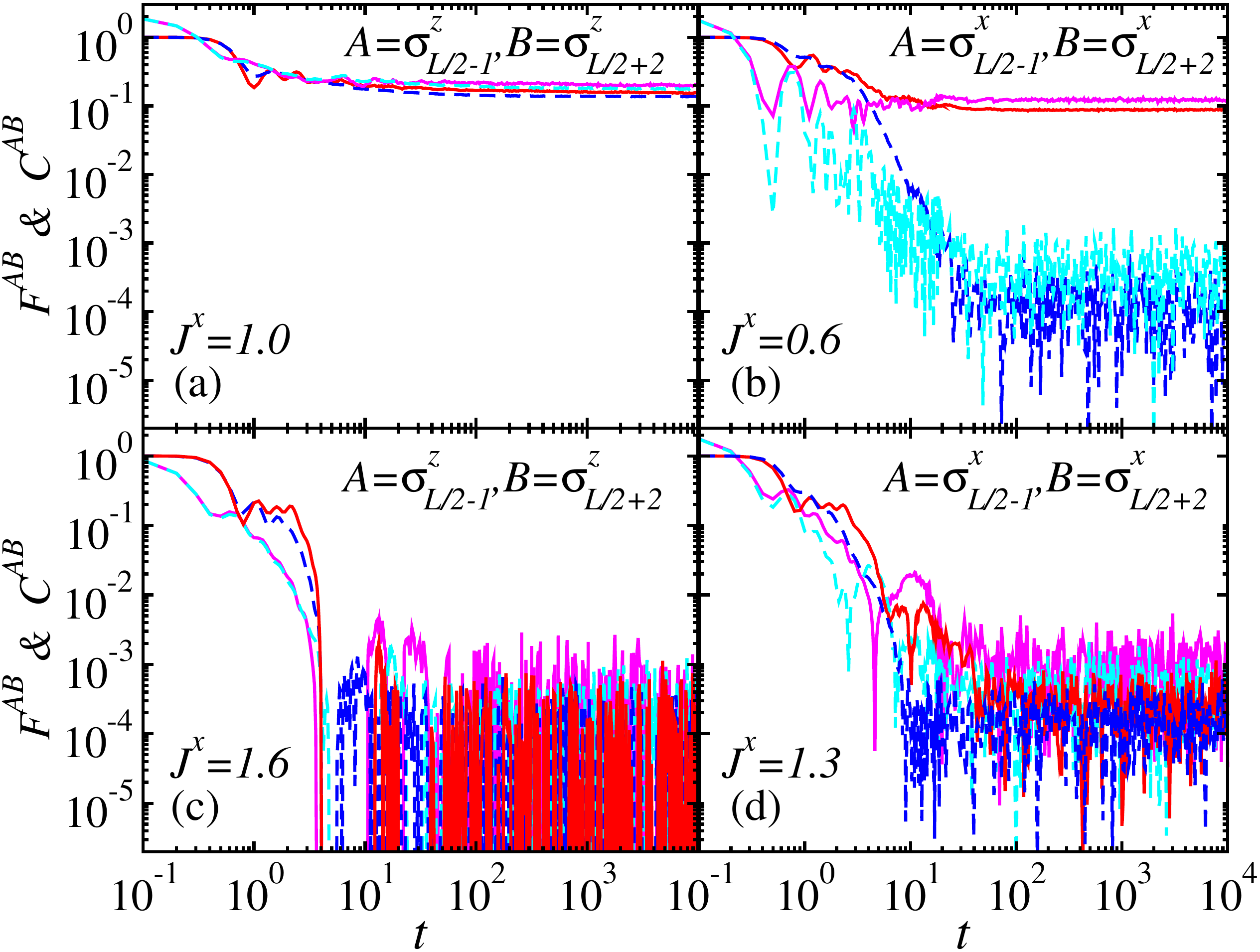}
\caption{Relaxation dynamics of $F^{AB}(t)$ and $C^{AB}(t)$ for the XYZ chain with $J^y=1.0$, $J^z=1.5$, and $L=14$. The solid lines correspond to the integrable model ($\Lambda = 0$) and the dashed lines to the nonintegrable one ($\Lambda = 0.2$). For the integrable (nonintegrable) model, the red (blue) line is for $F^{AB}(t)$ and the pink (cyan) line for $C^{AB}(t)$.}
\label{fig:fig2}
\end{figure}

In Fig.~\ref{fig:fig2}, we compare the dynamics of the OTOCs [$F^{AB}(t)$] and of the two-time correlators [$C^{AB}(t)$] for the integrable (solid lines) and nonintegrable (dashed lines) models. As expected from the analysis in Sec.~\ref{sec:otoc}, we observe a remarkable qualitative (and quantitative in the cases of slow dynamics) resemblance in the relaxation dynamics of the OTOCs and the two-time correlators.

In Figs.~\ref{fig:fig2}(a) and~\ref{fig:fig2}(c), $\hat A=\sz_{L/2-1}$ and $\hat B=\sz_{L/2+2}$ for $J^x=1.0$ and $J^x=1.6$, respectively. At the \blue{$U(1)$ points} considered in Fig.~\ref{fig:fig2}(a), the decay of the OTOC and of the two-time correlator are slow for the integrable and nonintegrable models, as expected. This contrasts with the fast relaxation seen in Fig.~\ref{fig:fig2}(c) for both models and quantities for $J^x=1.6$ [which is away from the peaks in Fig.~\ref{fig:fig1}(a)], so that $\sz_{L/2-1}$ and $\sz_{L/2+2}$ have no diagonal matrix elements. The saturation and oscillations at long times in all panels of Fig.~\ref{fig:fig2} are due to finite-size effects.

In Figs.~\ref{fig:fig2}(b) and~\ref{fig:fig2}(d), we show results for $A= \sx_{L/2-1} $ and $B=\sx_{L/2+2}$ for $J^x=0.6$ and $J^x=1.3$, respectively. \blue{For the integrable case at the SUSY point ($J^x=0.6$, $\Lambda=0$)}, Fig.~\ref{fig:fig2}(b) exhibits the expected slow decay of $F^{AB}(t)$ and $C^{AB}(t)$. \blue{Because of the proximity to the SUSY point, the behaviors of those quantities for $\Lambda=0.2$ in Fig.~\ref{fig:fig2}(b) resemble that at the SUSY point} for short times ($t\lesssim 2$), but the relaxation is fast for $t\gtrsim 2$. In Fig.~\ref{fig:fig2}(d), where $J^x=1.3$, none of the models presents any special symmetry, so both the OTOC and the two-time correlator exhibit a fast (exponential-like) decay, similar to the one observed in Fig.~\ref{fig:fig2}(c).

\subsection{Relaxation dynamics and off-diagonal elements of local operators}\label{sec:otoc_twotime}

The resemblance between the dynamics of the OTOC and the two-time correlator in the presence of symmetries seen in Sec.~\ref{sec:numdiag} implies that the functions $f^A(E_\infty, \omega)$ that characterize the off-diagonal matrix elements of $\sa_l$ can be used to gain a qualitatively understanding of the OTOC decay in interacting integrable and nonintegrable many-body quantum models. $f^A(E_\infty, \omega)$ is expected to exhibit different low-frequency behaviors depending on whether the OTOC decays slowly or fast.

\begin{figure}[!t]
\includegraphics[width=\columnwidth]{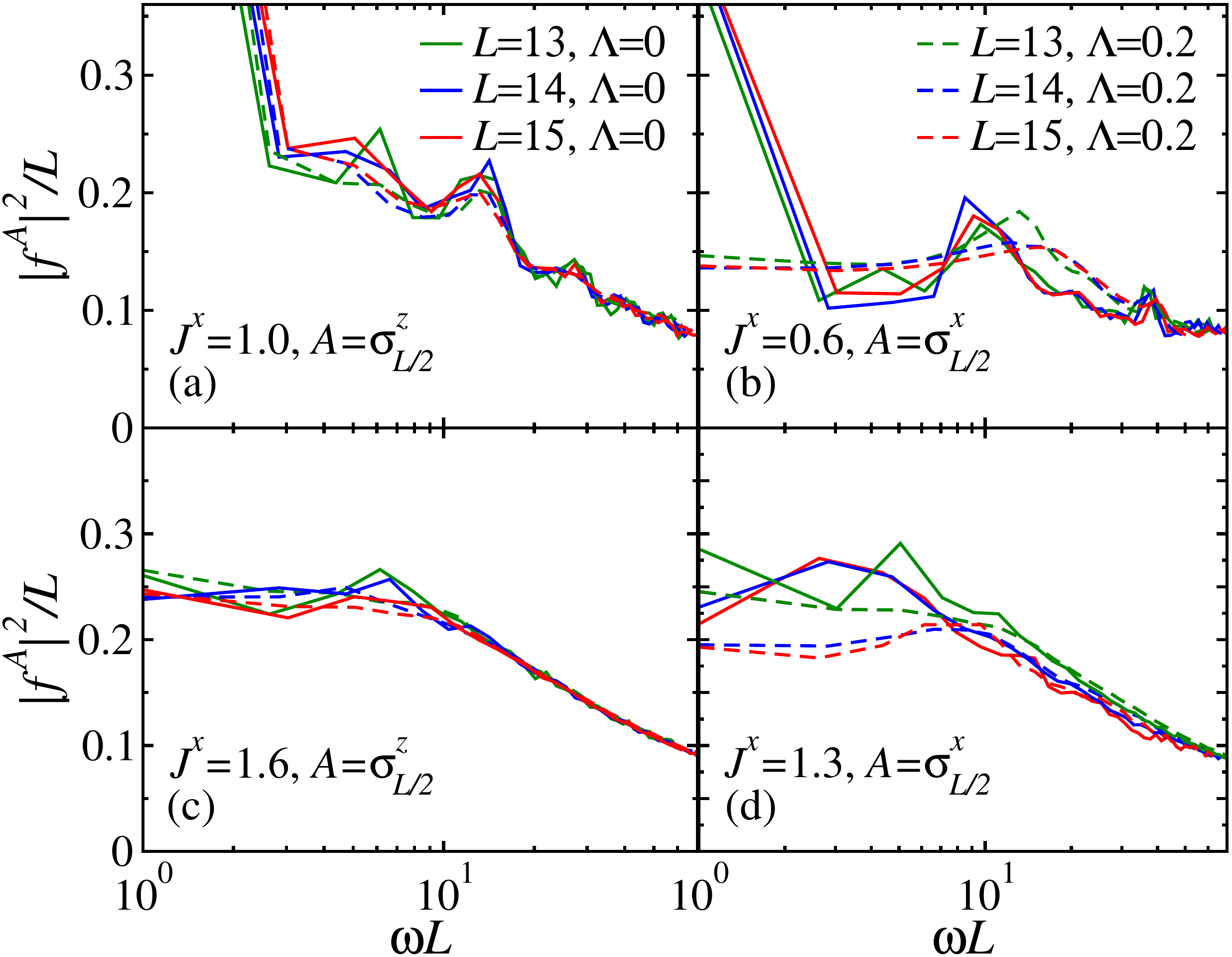}
\caption{Plots of $\left|f^{\sz_{L/2}}(E_\infty, \omega)\right|^2/L$ [(a) and (c)] and $\left|f^{\sx_{L/2}}(E_\infty, \omega)\right|^2/L$ [(b) and (d)] as functions of $\omega L$ for the XYZ chain with $J^y=1.0$ and $J^z=1.5$. Solid (dashed) lines are for the integrable (nonintegrable) chain.}
\label{fig:fig3}
\end{figure}

Following the discussion in Sec.~\ref{sec:otoc_var}, we can match the relaxation dynamics of the OTOCs in Fig.~\ref{fig:fig2} with the behavior of $\left|f^A(E_\infty, \omega)\right|^2$ in Fig.~\ref{fig:fig3}. The smooth function $\left|f^A(E_\infty, \omega)\right|^2$ is calculated from the variance $\text{Var}$ of the off-diagonal matrix elements of $A$ as $\left|f^A(E_\infty, \omega)\right|^2\approx\mathcal{V} \text{Var}(A_{\alpha \beta})$ \cite{LeBlond2019, LeBlond2020}. Specifically, we average over all off-diagonal matrix elements $A_{\alpha \beta}$ in a frequency interval $\Delta\omega=0.01$ centered at points separated by $\delta\omega=0.002$. In Fig.~\ref{fig:fig3}, we show results for three different system sizes, as indicated in the legend.

We find a one-to-one correspondence between Figs.~\ref{fig:fig2} and~\ref{fig:fig3}. The OTOC decays slowly [Figs.~\ref{fig:fig2}(a) and~\ref{fig:fig2}(b) at the \blue{SUSY point}] when $\left|f^A(E_\infty, \omega)\right|^2$ exhibits a peak at low frequencies [Figs.~\ref{fig:fig3}(a) and~\ref{fig:fig3}(b) at the \blue{SUSY point}], but it decays fast [Figs.~\ref{fig:fig2}(c) and~\ref{fig:fig2}(d), and Fig.~\ref{fig:fig2}(b) for the nonintegrable model] when $\left|f^A(E_\infty, \omega)\right|^2$ exhibits a plateau at low frequencies [Figs.~\ref{fig:fig3}(c) and~\ref{fig:fig3}(d), and Fig.~\ref{fig:fig3}(b) for the nonintegrable model]. For the system sizes considered, one can see in Figs.~\ref{fig:fig3}(c) and~\ref{fig:fig3}(d) that $\left|f^A(E_\infty, \omega)\right|^2$ is smoother at low frequencies for the nonintegrable model than for the integrable one. Since the results for the integrable model suffer from stronger finite-size effects, we expect those differences to disappear with increasing the system size.

\section{Summary}
\label{sec:conclude}

The scrambling of quantum information as characterized by the relaxation dynamics of the OTOCs has been mainly investigated in nonintegrable models. Here, we report a study of OTOCs in interacting integrable and nonintegrable spin-$\tfrac{1}{2}$ XYZ chains, in regimes without a classical counterpart.

We show that the main factor determining the behavior of the OTOCs in those chains (for the timescales and chain sizes explored) is the overlap between its constituent operators and conserved quantities of the model, rather than integrability or lack thereof. If the overlap is nonzero, then the OTOCs decay slowly (algebraically-like) for both the integrable and nonintegrable chains, while if there is no overlap, then the OTOCs decay rapidly (exponentially-like) for both the integrable and the nonintegrable chains. For the model Hamiltonians and operators considered, nonzero overlaps are obtained only in the presence of $U(1)$ symmetry and supersymmetry. Our numerical results for the dynamics of the OTOCs can be understood, in a complementary manner, in terms of the behavior of the off-diagonal matrix elements of the OTOCs operators at low frequencies and of the scaling with the system size of the diagonal matrix elements.

Furthermore, our analytical results (confirmed numerically) also indicate that when the operators of interest overlap with the Hamiltonian (a conserved quantity under the dynamics studied), the infinite-time average of the two-time correlators in finite (but sufficiently large) chains determines the infinite-time average of the OTOCs. Hence, in the thermodynamic limit (via the Lieb-Robinson bound argument presented in Sec.~\ref{sec:otoc}), the slow relaxation of the OTOC can be understood to be a consequence of the slow relaxation of two-time correlators, all resulting from the presence of a conservation law and independent of the integrable or nonintegrable nature of the system.

\begin{acknowledgments}
L.F.S.~is supported by the United States National Science Foundation (NSF) Grant No.~DMR-1936006 and the MPS Simons Foundation Award ID: 678586. M.R.~is supported by the United States National Science Foundation (NSF) Grant No.~PHY-2012145. D.P.~acknowledges support from the Ministry of Education Singapore, under the Grant No.~MOE-T2EP50120-0019. The computational work for this article was partially performed on the National Supercomputing Centre, Singapore \cite{NSCC}. 

\end{acknowledgments}

\appendix

\section{Even vs odd chains}
\label{app:odd-even}

\begin{figure}[!h]
\includegraphics[width=\columnwidth]{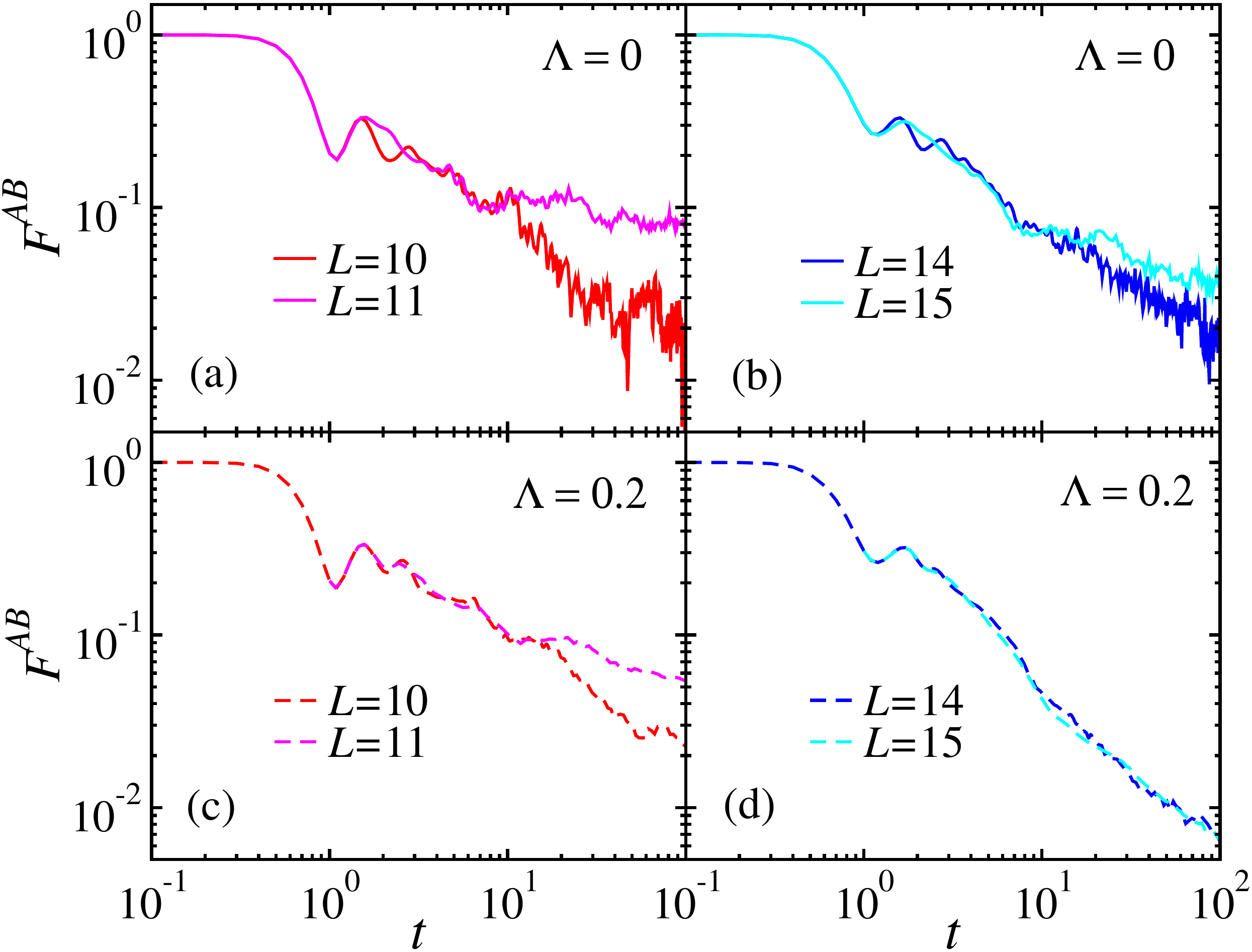}
\caption{Comparison of the relaxation dynamics of  $F^{\sz_{L/2-1} \sz_{L/2+2}}$ for different sizes of the XYZ chain with $J^x=0.8$, $J^y=1.0$, and $J^z=1.5$. Solid (dashed) lines are for the integrable (nonintegrable) chain.}
\label{fig:figS1}
\end{figure}

A comparison between the decay of the OTOCs for even and odd chains, away from the points with $U(1)$ symmetry or supersymmetry, is provided in Fig.~\ref{fig:figS1} for the integrable [Figs.~\ref{fig:figS1}(a) and~\ref{fig:figS1}(b)] and the nonintegrable [Figs.~\ref{fig:figS1}(c) and~\ref{fig:figS1}(d)] model. As the system size increases, one can see an increase of the time up to which the dynamics for the even and odd sizes coincide. This lends support to our conclusion that for sufficiently large system sizes our analysis based only on the study of even system sizes will also apply to odd system sizes.

\bibliography{final}

\end{document}